
\newif\iflanl
\openin 1 lanlmac
\ifeof 1 \lanlfalse \else \lanltrue \fi
\closein 1
\iflanl
    \input lanlmac
\else
    \message{[lanlmac not found - use harvmac instead}
    \input harvmac
\fi
\newif\ifhypertex
\ifx\hyperdef\UnDeFiNeD
    \hypertexfalse
    \message{[HYPERTEX MODE OFF}
    \def\hyperref#1#2#3#4{#4}
    \def\hyperdef#1#2#3#4{#4}
    \def\hypernoname{}
    \def\e@tf@ur#1{}
    \def\hth/#1#2#3#4#5#6#7{{\tt hep-th/#1#2#3#4#5#6#7}}
    \def\CERN{\address{CERN, Geneva, Switzerland}}
\else
    \hypertextrue
    \message{[HYPERTEX MODE ON}
  \def\hth/#1#2#3#4#5#6#7{
  {\tt hep-th/#1#2#3#4#5#6#7}}
\def\CERN{\address{

Theory Division, CERN, Geneva, Switzerland}}
\fi
\newif\ifdraft

\noblackbox
\catcode`\@=11
\newif\iffrontpage
\ifx\answ\bigans
\def\titleft{\titsm}
\magnification=1200\baselineskip=14pt plus 2pt minus 1pt
%
\advance\hoffset by-0.075truein
\advance\voffset by1.truecm
\hsize=6.15truein\vsize=600.truept\hsbody=\hsize\hstitle=\hsize
\else\let\lr=L
\def\titleft{\titla}
\magnification=1000\baselineskip=14pt plus 2pt minus 1pt
%
\hoffset=-0.75truein\voffset=-.0truein
\vsize=6.5truein
\hstitle=8.truein\hsbody=4.75truein
\fullhsize=10truein\hsize=\hsbody
\fi
\parskip=4pt plus 15pt minus 1pt
%
\newif\iffigureexists
\newif\ifepsfloaded
\def\epsfcheck{
\ifdraft
\input epsf\epsfloadedtrue
\else
  \openin 1 epsf
  \ifeof 1 \epsfloadedfalse \else \epsfloadedtrue \fi
  \closein 1
  \ifepsfloaded
    \input epsf
  \else
\immediate\write20{NO EPSF FILE --- FIGURES WILL BE IGNORED}
  \fi
\fi
\def\epsfcheck{}}
\def\checkex#1{
\ifdraft
\figureexistsfalse\immediate%
\write20{Draftmode: figure #1 not included}
\else\relax
    \ifepsfloaded \openin 1 #1
        \ifeof 1
           \figureexistsfalse
  \immediate\write20{FIGURE FILE #1 NOT FOUND}
        \else \figureexiststrue
        \fi \closein 1
    \else \figureexistsfalse
    \fi
\fi}
\def\missbox#1#2{$\vcenter{\hrule
\hbox{\vrule height#1\kern1.truein
\raise.5truein\hbox{#2} \kern1.truein \vrule} \hrule}$}
\def\lfig#1{
\let\labelflag=#1%
\def\numb@rone{#1}%
\ifx\labelflag\UnDeFiNeD%
{\xdef#1{\the\figno}%
\writedef{#1\leftbracket{\the\figno}}%
\global\advance\figno by1%
}\fi{\hyperref{}{figure}{{\numb@rone}}{Fig.{\numb@rone}}}}
\def\figinsert#1#2#3#4{
\epsfcheck\checkex{#4}%
\def\figsize{#3}%
\let\flag=#1\ifx\flag\UnDeFiNeD
{\xdef#1{\the\figno}%
\writedef{#1\leftbracket{\the\figno}}%
\global\advance\figno by1%
}\fi
\goodbreak\midinsert%
\iffigureexists
\centerline{\epsfysize\figsize\epsfbox{#4}}%
\else%
\vskip.05truein
  \ifepsfloaded
  \ifdraft
  \centerline{\missbox\figsize{Draftmode: #4 not included}}%
  \else
  \centerline{\missbox\figsize{#4 not found}}
  \fi
  \else
  \centerline{\missbox\figsize{epsf.tex not found}}
  \fi
\vskip.05truein
\fi%
{\smallskip%
\leftskip 4pc \rightskip 4pc%
\noindent\ninepoint\sl \baselineskip=11pt%
{\bf{\hyperdef\hypernoname{figure}{{#1}}{Fig.{#1}}}:~}#2%
\smallskip}\bigskip\endinsert%
}
%
\font\bigit=cmti10 scaled \magstep1

\font\titla=cmr10 scaled\magstep3
\font\tenmss=cmss10
\font\absmss=cmss10 scaled\magstep1

\newfam\mssfam
\font\footrm=cmr8  \font\footrms=cmr5
\font\footrmss=cmr5   \font\footi=cmmi8
\font\footis=cmmi5   \font\footiss=cmmi5
\font\footsy=cmsy8   \font\footsys=cmsy5
\font\footsyss=cmsy5   \font\footbf=cmbx8
\font\footmss=cmss8
\def\footfont{\def\rm{\fam0\footrm}
\textfont0=\footrm \scriptfont0=\footrms
\scriptscriptfont0=\footrmss
\textfont1=\footi \scriptfont1=\footis
\scriptscriptfont1=\footiss
\textfont2=\footsy \scriptfont2=\footsys
\scriptscriptfont2=\footsyss
\textfont\itfam=\footi \def\it{\fam\itfam\footi}
\textfont\mssfam=\footmss \def\mss{\fam\mssfam\footmss}
\textfont\bffam=\footbf \def\bf{\fam\bffam\footbf} \rm}
\def\tenpoint{\def\rm{\fam0\tenrm}
\textfont0=\tenrm \scriptfont0=\sevenrm
\scriptscriptfont0=\fiverm
\textfont1=\teni  \scriptfont1=\seveni
\scriptscriptfont1=\fivei
\textfont2=\tensy \scriptfont2=\sevensy
\scriptscriptfont2=\fivesy
\textfont\itfam=\tenit \def\it{\fam\itfam\tenit}
\textfont\mssfam=\tenmss \def\mss{\fam\mssfam\tenmss}
\textfont\bffam=\tenbf \def\bf{\fam\bffam\tenbf} \rm}
\ifx\answ\bigans\def\abstractfont{\tenpoint}\else
\def\abstractfont{\def\rm{\fam0\absrm}
\textfont0=\absrm \scriptfont0=\absrms
\scriptscriptfont0=\absrmss
\textfont1=\absi \scriptfont1=\absis
\scriptscriptfont1=\absiss
\textfont2=\abssy \scriptfont2=\abssys
\scriptscriptfont2=\abssyss
\textfont\itfam=\bigit \def\it{\fam\itfam\bigit}
\textfont\mssfam=\absmss \def\mss{\fam\mssfam\absmss}
\textfont\bffam=\absbf \def\bf{\fam\bffam\absbf}\rm}\fi
%
\def\f@@t{\baselineskip10pt\lineskip0pt\lineskiplimit0pt
\bgroup\aftergroup\@foot\let\next}
\setbox\strutbox=\hbox{\vrule height 8.pt depth 3.5pt width\z@}
\def\vfootnote#1{\insert\footins\bgroup
\baselineskip10pt\footfont
\interlinepenalty=\interfootnotelinepenalty
\floatingpenalty=20000
\splittopskip=\ht\strutbox \boxmaxdepth=\dp\strutbox
\leftskip=24pt \rightskip=\z@skip
\parindent=12pt \parfillskip=0pt plus 1fil
\spaceskip=\z@skip \xspaceskip=\z@skip
\Textindent{$#1$}\footstrut\futurelet\next\fo@t}
\def\Textindent#1{\noindent\llap{#1\enspace}\ignorespaces}
\def\foot{\global\advance\ftno by1%
\attach{\hyperref{}{footnote}{\the\ftno}{\footsymbolgen}}%
\vfootnote{\hyperdef\hypernoname{footnote}{\the\ftno}{\footsymbol}}}%
\def\footnote#1{\global\advance\ftno by1%
\attach{\hyperref{}{footnote}{\the\ftno}{#1}}%
\vfootnote{\hyperdef\hypernoname{footnote}{\the\ftno}{#1}}}%
\newcount\lastf@@t           \lastf@@t=-1
\newcount\footsymbolcount    \footsymbolcount=0
\global\newcount\ftno \global\ftno=0
\def\footsymbolgen{\relax\footsym
\global\lastf@@t=\pageno\footsymbol}
\def\footsym{\ifnum\footsymbolcount<0
\global\footsymbolcount=0\fi
{\iffrontpage \else \advance\lastf@@t by 1 \fi
\ifnum\lastf@@t<\pageno \global\footsymbolcount=0
\else \global\advance\footsymbolcount by 1 \fi }
\ifcase\footsymbolcount
\fd@f\dagger\or \fd@f\diamond\or \fd@f\ddagger\or
\fd@f\natural\or \fd@f\ast\or \fd@f\bullet\or
\fd@f\star\or \fd@f\nabla\else \fd@f\dagger
\global\footsymbolcount=0 \fi }
\def\fd@f#1{\xdef\footsymbol{#1}}
\def\space@ver#1{\let\@sf=\empty \ifmmode #1\else \ifhmode
\edef\@sf{\spacefactor=\the\spacefactor}
\unskip${}#1$\relax\fi\fi}
\def\attach#1{\space@ver{\strut^{\mkern 2mu #1}}\@sf}
%
\newif\ifnref
\def\rrr#1#2{\relax\ifnref\nref#1{#2}\else\ref#1{#2}\fi}
\def\ldf#1#2{\begingroup\obeylines
\gdef#1{\rrr{#1}{#2}}\endgroup\unskip}

\def\doubref#1#2{\refs{{#1},{#2}}}

\nreffalse
%

%
\def\eqn#1{\xdef #1{(\noexpand\hyperref{}%
{equation}{\secsym\the\meqno}%
{\secsym\the\meqno})}\eqno(\hyperdef\hypernoname{equation}%
{\secsym\the\meqno}{\secsym\the\meqno})\eqlabeL#1%
\writedef{#1\leftbracket#1}\global\advance\meqno by1}
\def\eqnalign#1{\xdef #1{\noexpand\hyperref{}{equation}%
{\secsym\the\meqno}{(\secsym\the\meqno)}}%
\writedef{#1\leftbracket#1}%
\hyperdef\hypernoname{equation}%
{\secsym\the\meqno}{\e@tf@ur#1}\eqlabeL{#1}%
\global\advance\meqno by1}
\def\eqnalign#1{\xdef #1{(\secsym\the\meqno)}
\writedef{#1\leftbracket#1}%
\global\advance\meqno by1 #1\eqlabeL{#1}}
%

%
\def\chap#1{\newsec{#1}}
\def\chapter#1{\chap{#1}}
\def\sect#1{\subsec{#1}}
\def\section#1{\sect{#1}}
\def\\{\ifnum\lastpenalty=-10000\relax
\else\hfil\penalty-10000\fi\ignorespaces}
\def\note#1{\leavevmode%
\edef\@@marginsf{\spacefactor=\the\spacefactor\relax}%
\ifdraft\strut\vadjust{%
\hbox to0pt{\hskip\hsize%
\ifx\answ\bigans\hskip.1in\else\hskip .1in\fi%
\vbox to0pt{\vskip-\dp
\strutbox\sevenbf\baselineskip=8pt plus 1pt minus 1pt%
\ifx\answ\bigans\hsize=.7in\else\hsize=.35in\fi%
\tolerance=5000 \hbadness=5000%
\leftskip=0pt \rightskip=0pt \everypar={}%
\raggedright\parskip=0pt \parindent=0pt%
\vskip-\ht\strutbox\noindent\strut#1\par%
\vss}\hss}}\fi\@@marginsf\kern-.01cm}
\def\titlepage{%
\frontpagetrue\nopagenumbers\abstractfont%
\hsize=\hstitle\rightline{\vbox{\baselineskip=10pt%
{\abstractfont\pubnum}}}\pageno=0}
\frontpagefalse
\def\pubnum{}
\def\pdate{\number\month/\number\yearltd}
\def\makefootline{\iffrontpage\vskip .27truein
\line{\the\footline}
\vskip -.1truein\leftline{\vbox{\baselineskip=10pt%
{\abstractfont\pdate}}}
\else\vskip.5cm\line{\hss \tenrm $-$ \folio\ $-$ \hss}\fi}
\def\title#1{\vskip .7truecm\titlestyle{\titleft #1}}
\def\titlestyle#1{\par\begingroup \interlinepenalty=9999
\leftskip=0.02\hsize plus 0.23\hsize minus 0.02\hsize
\rightskip=\leftskip \parfillskip=0pt
\hyphenpenalty=9000 \exhyphenpenalty=9000
\tolerance=9999 \pretolerance=9000
\spaceskip=0.333em \xspaceskip=0.5em
\noindent #1\par\endgroup }
\def\autskip{\ifx\answ\bigans\vskip.5truecm\else\vskip.1cm\fi}
\def\author#1{\vskip .7in \centerline{#1}}

\def\address#1{\ifx\answ\bigans\vskip.2truecm
\else\vskip.1cm\fi{\it \centerline{#1}}}
\def\abstract#1{
\vskip .5in\vfil\centerline
{\bf Abstract}\penalty1000
{{\smallskip\ifx\answ\bigans\leftskip 2pc \rightskip 2pc
\else\leftskip 5pc \rightskip 5pc\fi
\noindent\abstractfont \baselineskip=12pt
{#1} \smallskip}}
\penalty-1000}
\def\endpage{\tenpoint\supereject\global\hsize=\hsbody%
\frontpagefalse\footline={\hss\tenrm\folio\hss}}
\def\ack{\goodbreak\vskip2.cm\centerline{{\bf Acknowledgements}}}
\def\app#1{\goodbreak\vskip2.cm\centerline{{\bf Appendix: #1}}}
%
\def\bfone{\relax{\rm 1\kern-.35em 1}}
\def\inbar{\vrule height1.5ex width.4pt depth0pt}
\def\IC{\relax\,\hbox{$\inbar\kern-.3em{\mss C}$}}
\def\ID{\relax{\rm I\kern-.18em D}}
\def\IF{\relax{\rm I\kern-.18em F}}
\def\IH{\relax{\rm I\kern-.18em H}}
\def\II{\relax{\rm I\kern-.17em I}}
\def\IN{\relax{\rm I\kern-.18em N}}
\def\IP{\relax{\rm I\kern-.18em P}}
\def\IQ{\relax\,\hbox{$\inbar\kern-.3em{\rm Q}$}}
\def\IR{\relax{\rm I\kern-.18em R}}
\font\cmss=cmss10 \font\cmsss=cmss10 at 7pt
\def\ZZ{\relax\ifmmode\mathchoice
{\hbox{\cmss Z\kern-.4em Z}}{\hbox{\cmss Z\kern-.4em Z}}
{\lower.9pt\hbox{\cmsss Z\kern-.4em Z}}
{\lower1.2pt\hbox{\cmsss Z\kern-.4em Z}}\else{\cmss Z\kern-.4em
Z}\fi}

\def\G{\Gamma} \def\l{\lambda}

\def\cF{{\cal F}} 
 
 \def\cK{{\cal K}}

\def\nup#1({Nucl.\ Phys.\ $\us {B#1}$\ (}
\def\plt#1({Phys.\ Lett.\ $\us  {#1}$\ (}
\def\cmp#1({Comm.\ Math.\ Phys.\ $\us  {#1}$\ (}
\def\prp#1({Phys.\ Rep.\ $\us  {#1}$\ (}
\def\prl#1({Phys.\ Rev.\ Lett.\ $\us  {#1}$\ (}
\def\prv#1({Phys.\ Rev.\ $\us  {#1}$\ (}
\def\mpl#1({Mod.\ Phys.\ Let.\ $\us  {A#1}$\ (}
\def\ijmp#1({Int.\ J.\ Mod.\ Phys.\ $\us{A#1}$\ (}
\def\tit#1|{{\it #1},\ }
%

%

\def\ni{\noindent}
\def\tilde{\widetilde}
\def\bar{\overline}
\def\us#1{\underline{#1}}

\def\Coe#1.#2.{{#1\over #2}}
\def\coeff#1#2{\relax{\textstyle {#1 \over #2}}\displaystyle}
\def\coe#1.#2.{\relax{\textstyle {#1 \over #2}}\displaystyle}

\def\to{\rightarrow}
\def\notin{\hbox{{$\in$}\kern-.51em\hbox{/}}}

\def\del{\partial}

\def\nex#1{$N\!=\!#1$}

\catcode`\@=12

\def\cF{{\cal F}}

\def\tx    {\textsty\defle}

\def\xb   {\bar x}
\def\yb   {\bar y}
\def\zb   {\bar z}

\def\th  {\theta}

\def\tz{\theta_z}
\def\ty{\theta_y}
\def\tx{\theta_x}

\def\cicy#1(#2|#3)#4{\left(\matrix{#2}\right|\!\!
                     \left|\matrix{#3}\right)^{{#4}}_{#1}}

%
\ldf\KV{S.\ Kachru and C.\ Vafa, {\it Exact Results
for $N=2$ Compactifications of the Heterotic String}, preprint
HUTP-95/A016, \hth/9505105.}
\ldf\HKTYI{S.\ Hosono, A.\ Klemm, S.\ Theisen and S.-T. Yau,
\cmp167(1995) 301, \hth/9308122.}
\ldf\HKTYII{S.\ Hosono, A.\ Klemm, S.\ Theisen and S.-T. Yau,
\nup433(1995) 501, \hth/9406055.}
\ldf\SW{N.\ Seiberg and E.\ Witten, \nup426(1994) 19, \hth/9407087;
\nup431(1994) 484, \hth/9408099.}
\ldf\KLTYa{A.\ Klemm, W.\ Lerche, S.\ Theisen and S.\ Yankielowicz,
\plt344(1995) 169, \hth/9411048.}
\ldf\BKK{P.\ Berglund, S.\ Katz and A.\ Klemm, {\it
Mirror Symmetry and the Moduli Space for
Generic Hypersurfaces in Toric Varieties},
CERN-TH-7528/95, IASSNS-HEP-94/106, \hth/9506091.}
\ldf\LY{B.\ H.\ Lian and S.-T.\ Yau, {\it
Arithmetic Properties of the Mirror Map and Quantum Coupling},
preprint
\hth/9411234}
\ldf\dWKLL{B.\ de Wit, V.\ Kaplunovsky, J.\ Louis and D.\
L\"{u}st, {\it Perturbative Couplings of Vector Multiplets in $N=2$
Heterotic String Vacua}, \hth/9504006.}
\ldf\AFGNT{I.\ Antoniadis, S.\ Ferrara,
E.\ Gava, K.S.\ Narain and T.R.\ Taylor, {\it Perturbative
Prepotential and Monodromies in N=2 Heterotic Superstring},
\hth/9504034.}
\ldf\SergioEtAl{A.\ Ceresole, R.\ D'Auria and S.\ Ferrara,
\plt339(1994) 71, \hth/9408036; A.\ Ceresole, R.\ D'Auria, S.\
Ferrara and A.\ Van Proeyen, {\it On Electromagnetic Duality in
Locally Supersymmetric N=2 Yang--Mills Theory}, preprint
CERN-TH.7510/94, POLFIS-TH. 08/94, UCLA 94/TEP/45, KUL-TF-94/44,
\hth/9412200; {\it Duality Transformations in Supersymmetric
Yang-Mills Theories coupled to Supergravity}, preprint CERN-TH
7547/94, POLFIS-TH. 01/95, UCLA 94/TEP/45, KUL-TF-95/4, \hth/9502072;
M. Billo', A. Ceresole, R. D'Auria, S. Ferrara, P. Fre', T. Regge, P.
Soriani and A. Van Proeyen, {\it A Search for Non-Perturbative
Dualities of Local N=2 Yang--Mills Theories from Calabi--Yau
Threefolds}, preprint SISSA 64/95/EP, POLFIS-TH 07/95, CERN-TH
95/140, IFUM 508/FT, KUL-TF-95/18, UCLA/95/TEP/19, \hth/9506075.}
\ldf\FHSV{S. Ferrara, J. A. Harvey, A. Strominger and C. Vafa, {\it
Second-Quantized Mirror Symmetry}, preprint EFI-95-26, \hth/9505162.}
\ldf\Andy{A.\ Strominger, {\it Massless Black Holes and Conifolds in
String Theory}, ITP St.\ Barbara preprint, \hth/9504090.}
\ldf\LSW{W.\ Lerche, D.\ Smit and N.\ Warner, \nup372 (1992) 87,
\hth/9108013.}
\ldf\COFKM{P.\ Candelas, X.\ de la Ossa, A.\ Font, S.\ Katz and D.\
Morrison,
\nup416 (1994) 481, \hth/9308083.}
\ldf\PrivComm{S.\ Theisen and C.\ Vafa, private communication.}
\ldf\CN{J.\ Conway and S.\ Norton, Bull.\ Math.\ Soc.\ 11 (1979)
308.}
\ldf\BPSmass{See the third reference in \SergioEtAl; G.\ Cardoso, L.\
Iba\~nez, D.\ L\"ust and T.\ Mohaupt, in preparation.}
\ldf\Yone{T.\ Yonemura, Tohuku Math.\ J.\ 42 (1990) 351.}
%
\def\pubnum{
\hbox{CERN-TH/95-165}
\hbox{hep-th/9506112}}
\def\pdate{}
\titlepage
\vskip2.cm
\title
{K3--Fibrations and Heterotic-Type II String Duality}
\vskip.5cm\autskip
\author{A.\ Klemm, W.\ Lerche and  P.\ Mayr}
\CERN
\vskip .5truecm
\vskip-1.7truecm
\vskip-1.2truecm
\abstract{
We analyze the map between heterotic and type II N=2 supersymmetric
string theories for certain two and three moduli examples found by
Kachru and Vafa. The appearance of elliptic j-functions can be traced
back to specializations of the Picard-Fuchs equations to systems for
$K_3$ surfaces. For the three-moduli example we write the mirror maps
and Yukawa couplings in the weak coupling limit in terms of
j-functions; the expressions agree with those obtained in
perturbative calculations in the heterotic string in an impressive
way. We also discuss symmetries of the world-sheet instanton numbers
in the type II theory, and interpret them in terms of S--duality of
the non-perturbative heterotic string.}
\vfil
\vskip 1.cm
\ni {CERN-TH/95-165}\hfill\break
\ni June  1995
\endpage
\baselineskip=14pt plus 2pt minus 1pt
\sequentialequations

\chapter{Introduction}

In a very interesting recent paper \KV, Kachru and Vafa provided
concrete evidence of the conjecture that the exact non-perturbative
behavior of the heterotic string compactified on $K_3\times T_2$ is
governed by certain Calabi-Yau (CY) manifolds \SergioEtAl, and can
effectively be described in terms of type II strings \KLTYa.

In particular, there are examples \doubref\KV\FHSV\ of CY's that,
when taken as background of type II theories, lead to prepotentials
that reproduce certain perturbative corrections of the heterotic
theory in the weak string coupling limit (for non-zero coupling, one
expects new stringy non-perturbative phenomena \Andy\ to become
visible, analogous to rigid N=2 Yang-Mills theory \SW). It is known
from explicit heterotic string computations \doubref\dWKLL\AFGNT\
that these corrections are given in terms of elliptic $j$-functions
in the $T_2$ moduli. That precisely such combinations of
$j$-functions really do appear \COFKM\ in the moduli spaces of
certain CY manifolds, is highly suggestive, and at first rather
surprising.

One of the purposes of this letter is to gain insight in the origin
of such modular functions in the moduli spaces of certain
Calabi-Yau's. We will show that this can be very simply understood in
terms of specializations of Picard-Fuchs equations, and more
abstractly in terms of CY manifolds being elliptic or $K_3$
fibrations. This understanding opens up the possibility of a more
systematic construction of CY manifolds that describe the exact
quantum theory of \nex2 supersymmetric heterotic strings. In
particular, it also allows us to perform further non-trivial checks
on the original examples of Kachru and Vafa, by explicitly writing
certain ``Yukawa couplings''\foot{What we mean are the triple
derivatives of the prepotential, which have in the present context,
strictly speaking, the interpretation of anomalous magnetic moments.}
in terms of $j$-functions.

We will also briefly investigate the symmetry structure of certain
models, linking the symmetries of the CY instanton expansion to the
perturbative and non-perturbative $T$- and $S$-dualities of the
quantum heterotic string. In particular, we find evidence that in
some models there is a symmetry of exchanging the heterotic dilaton
$S$ with a target space moduli field, $T$.

\chapter{Modular properties of certain Calabi-Yau moduli sub-spaces}

The appearance of $j$-functions is the key for making the
relationship of heterotic strings compactified on $K_3\times T_2$
with type II compactifications evident \KV. Many of the examples of
``dual'' Calabi-Yau threefolds in \doubref\HKTYI\HKTYII\ are actually
elliptic fibrations over rational surfaces, or $K_3$ fibrations over
rational curves. In this section we show how these properties lead to the
crucial modular properties of the mirror map (and Yukawa couplings)
in the weak string coupling domain.

In fact, $K_3$ fibrations are of natural interest for the conjectured
duality between heterotic and type II compactifications, because they
automatically give rise to prepotentials in the large complex
structure limit of the form: $\cF=s Q(t) + C(t)$ (where $Q$ is
quadratic polynomial and $C$ is a cubic polynomial). That is, $s$ is
a good candidate for the heterotic dilaton.

Specifically, consider the model
$X_{12}(1,1,2,2,6)^{-252}_{2}$,\foot{We use the notation of
\HKTYI, e.g. $X_{d_1,\ldots, d_r} (w_1,\ldots,w_n)^{\chi}_{h_{1,1}}$ is a
complete
intersection (hypersurface) of multi-degree $d_1,\ldots,d_r$ in
weighted projective space $\IP^{n-1}(w_1,\ldots,w_n)$ (if all
$w_i=1$ we omit them) with Euler number $\chi$ and Betti number $h_{1,1}$.}
which is the first
of the examples discussed in \KV. The defining polynomial is
$$
p(x)\ =\
{x_1}^{12}+{x_2}^{12}+{x_3}^{6}+{x_4}^{6}+{x_5}^{2}
-12\psi\, x_1x_2x_3x_4x_5 - 2\phi\,{x_1}^{6}{x_2}^{6}\ ,
\eqn\firstex
$$
and the weak coupling limit was identified in \KV\ with $\yb={1\over
\phi^2} \rightarrow 0$ and $\xb=- {2 \phi\over (12 \psi^2)^3}$ finite.
In terms of these variables, the Picard-Fuchs operators look
($\tx\equiv x\del_x$, etc.):
$$
\eqalign{
{\cal D}_1 &=\tx^2\,(\tx-2\,\ty)-
8\, x \,(6\,\tx+5)\,(6\,\tx+3)\,(6\,\tx+1)\cr
{\cal D}_2&=\ty^2- y\,(2\,\ty-\tx+1)\,(2\,\ty-\tx)\ .}
\eqn\PFoper
$$
One way of understanding why a modular function appears in the
$y\to0$ limit is via the following two steps. First, the surface
\firstex\ is a $K_3$ fibration \COFKM\ in that there is a linear
system $|L|$ generated by the polynomial of degree one, whose
divisors are described after the substitution $x_1 =\lambda x_2$ and
the single-valued variable change $\tilde x_1=x_1^2$ as following
family of degree $12$ $K_3$ hypersurfaces in $\IP^{(1,1,1,3)}$:
$$
\cK\,:\ \ (1+\l^{12}-2 \phi\l^6) {\tilde x_1}^6+x_3^6+x_4^6+x_5^2-12
\psi \lambda \tilde x_1 x_3 x_4 x_5=0
\eqn\firstKIII
$$
As divisors in $|L|$ are disjoint, $|L|\cdot|L|=0$ holds and thus the
cubic intersection form has indeed the desired property. Moreover,
taking the above limit $\phi\rightarrow \infty$, $\psi \sim
\phi^{1/6} \tilde \psi$ and $\l \sim \phi^{-1/6}$ all terms in
\firstKIII\ stay finite, and $x=- {2 \over (12 \tilde \psi^2)^3}$
can be identified as the canonical one-parameter deformation of
$\cK$.

Second, there is strong evidence that one-modulus deformations of
$K_3$ surfaces are intrinsically related to modular functions \LY.
That is, it was observed in \LSW\ that the $W_3$-invariant of any
single-modulus Picard-Fuchs operator of $K_3$ vanishes, and since
$W_2$ transforms under coordinate changes $z\rightarrow \zeta(z)$ as
$W_2\rightarrow W_2+ \{\zeta,z\}$, it is possible to rewrite the PF
operator, after gauging away $W_2$, as ${\cal D}=\partial_t^3$. In
order to implement this gauging, one needs to solve a Schwarzian
differential equation of the form
$$
\big\{x,t_x\big\}\ =\ 2 Q(x) \,(\del_{t_x} x(t_x))^2\
\eqn\schwarz
$$
for some $Q(x)$. Its solution $t_x(x)$ is given by a triangle
function $s(x)$, whose inverse yields a modular function that is
automatically associated with some discrete subgroup of $PSL(2,\IR)$.
(Equivalently, the mirror map $x(q_x)$, where $q_x\equiv e^{2 \pi i
t_x}$, is given by the ratio of two independent solutions of the
associated PF-system.)

It was observed in \LY\ that in all examples investigated so far this
subgroup is given by a subgroup of the modular group $SL(2,\ZZ)$
(possibly together with some extra Atkin-Lehner involutions), and the
authors conjectured this to be true for general one-modulus
deformations of $K_3$ arising from orbifold constructions.\foot{More
precisely, they conjectured the mirror maps to be given by Thompson
series, which have an intrinsic relationship to modular functions and
to the representation theory of the Convay-Norton groups.} In the
present example, $Q(x)={1-1968x+2654208x^2\over4x^2(1728x-1)^2}$,
$t_x=s(\coeff12,\coeff13,0;j(q_x))$, and this leads indeed to
$x=1/j(q_x)$ (this feature of the mirror map for vanishing $y$ was
noticed first in \COFKM).

The occurrence of this kind of specialization to $K_3$ surfaces, with
similar modular properties, is actually ubiquitous in the class of
complete intersection (hypersurface) CY spaces. Consider, for
example, the families
$$
\eqalign{
A:&\,\,\, X_8(1,1,2,2,2)^{-86}_2\cr
B:&\,\,\, X_{6,4}(1,1,2,2,2,2)^{-132}_2\cr
C:&\,\,\, X_{4,4,4}(1,1,2,2,2,2,2)^{-112}_2\,}
\eqn\ABC
$$
and their associated PF systems, whose relevant parts are
$$
\eqalign{
A:&\,\,\,  {\cal D}_1=\tx^2 (2 \ty-\tx)+4 x(4 \tx+3) (4\tx+2)
(4\tx+1) \cr
B:&\,\,\,  {\cal D}_1=\tx^2 (2 \ty-\tx)+6 x(2 \tx+1) (3\tx+2)
(3\tx+1) \cr
C:&\,\,\,  {\cal D}_1=\tx^2 (2 \ty-\tx)+8 x (2 \tx+1)^3
\ .}\eqn\ABCpf
$$
Together with the first example \firstex, these examples represent
selected one-modulus $K_3$ fibrations, and the PF operators
\PFoper,\ABCpf\ effectively reduce under $y\to0$ to the following
list of $K_3$ operators \LY:\foot
{$\G_0(N)_+$ denotes a group that in general includes certain
Atkin-Lehner involutions besides $\G_0(N)$; see \CN\ for details.}

$$
\vbox{\offinterlineskip\tabskip=0pt
\halign{\strut\vrule#
&\hfil~$#$
&\vrule#&
{}~~$#$~~\hfil
&~~$#$~~\hfil
&~~$#$~~\hfil
&\vrule#\cr
\noalign{\hrule}
& &&K_3{\rm  family} & {\rm PF\,\, operator} &
{\rm  mod.\ group} & \cr
\noalign{\hrule}
& {\cal K}_{{\phantom A}} &&X_{6}(1,1,1,3)
& \th^3 - 8 x(6 \th+5)(6 \th+3) (6 \th + 1)
& \Gamma_0(1)\equiv\Gamma
&\cr
&{\cal K}_A &&X_{4}
&
\th^3 - 4 x (4 \th +3)(4 \th +2) (4 \th + 1)
&\Gamma_0(2)_+
& \cr
&{\cal K}_B &&X_{3,2}
& \th^3 - 6 x (2 \th +1) (3 \th + 2) (3\th + 1)
&\Gamma_0(3)_+
& \cr
&{\cal K}_C &&X_{2,2,2}
& \th^3 -  8 x (2 \th +1)^3
&\Gamma_0(4)_+& \cr
\noalign{\hrule}}
\hrule}$$

Model $A$ was briefly discussed in \KV, where it was conjectured that
the relevant modular group should be given by an extension of some
$\G_0(2^k)$; from the table we can infer that this is indeed true,
with $k=1$. The commensurability relations of the $K_3$ mirror maps
$x(q_x)$ with the $j$-function were explicitly worked out in \LY:
$$
\eqalign{
{\cal K} & :\, P(j,x)= 1-j x =0\,,\cr
{\cal K}_A & :\, P(j,x)= 1 + 432 x - j x + 62208 x^2  +
              207 j x^2  + 2985984 x^3
\cr&\qquad\qquad\qquad
 - 3456 j x^3  + j^2 x^3=0\ ,\qquad {\rm etc.}
}$$
For the first model we immediately recover $x=1/j(q_x)$.
Similarly, the mirror maps for the models $A,B,C$, when written in
the form $1/x(q_x)-c$ (with $c=104,42,24$), specialize to the
Hauptmodul of $\Gamma_0(N)$ for $N=2,3,4$, while
$y(q_x=0,q_y)={q_y\over (1+q_y)^2}$. They are given by certain
Thompson series \CN, which can be written in terms of modular
functions as follows:
$$
\eqalign{
&A:\ x(q_x,q_y=0)\ =\  {16 (\eta(q_x)\eta(q_x^2))^8\over
(\vartheta_3^4+\vartheta_0^4)^4} \cr
&B:\ x(q_x,q_y=0)\ =\
\Big(
{\eta^{12}(q_x)\over\eta^{12}({q_x}^3)}
+729{\eta^{12}({q_x}^3)\over\eta^{12}({q_x})}
+54\Big)^{-1} \cr
&C:\ x(q_x,q_y=0)\ =\
{\eta^{24}(q_x)\eta^{24}({q_x}^4)\over\eta^{48}({q_x}^2)}\cr
}$$
These expressions might be useful for further checks on the
conjectured heterotic-type II string duality.

A simple generalization of \firstex\ and example $A$ would be to take
$K_3$ hypersurfaces in weighted $\IP^3$ of the form
$X_d(1,w_3,w_4,w_5)$, and consider as CY the $K_3$ fibration $X_{2
d}(1,1,2 w_3,2 w_4,2 w_5)$\foot{For $K_3$ fibrations one can always
choose a basis s.t. $\cF=s Q(t) + C(t)$ and $\int c_2 \wedge s=24$.
Conversely given this topological data one has still to find the suitable
projection map. For hypersurfaces in toric varieties there seem to be
the combinatorical condition, that the $K_3$ polyhedron
is embedded in the Calabi-Yau polyhedron.}. From the examples
we have checked, it appears that the discriminant
naturally is of the form $\triangle = \triangle(K_3)^2+\ ...$ , where
the dots denote terms which vanish in an appropriate (weak coupling) limit.
There are $95$
transversal families of such
$K_3$ surfaces \Yone; $31$ of them with $w_1=1$ give rise to transversal
Calabi-Yau configurations  and are listed in the Appendix.
It would be very interesting to investigate whether these
CY manifolds describe non-perturbative quantum heterotic strings.

In fact, the surface $X_{24}(1,1,2,8,12)^{-480}_{3}$, which was
studied too in \KV, is precisely of this type. The defining
polynomial is given by
$$
p = x_1^2 + x_2^3 + x_3^{12} + x_4^{24} + x_5^{24} -
12 \psi_0 x_1x_2x_3x_4x_5-2\psi_1 (x_3x_4x_5)^6-
\psi_2(x_4x_5)^{12}\,,
\eqn\npolI
$$
and variables that are appropriate near the point of maximal
unipotent monodromy in the large complex structure limit are: $x =
-{2\psi_1\over 1728^2\psi_0^6}$, $y = {1\over \psi_2^2}$, $z =
-{\psi_2\over 4\psi_1^2}$. For $y\rightarrow 0$, the PF-system
$$
\eqalign{
{\cal D}_1 &=\tx\,(\tx-2\,\tz)-12\, x\,(6\,\tx+5)\,(6\,\tx+1) \cr
{\cal D}_2 &=\ty^2- y\, (2\,\ty-\tz+1)\,(2\,\ty-\tz) \cr
{\cal D}_3 &=\tz \,(\tz-2\ty)- z (2\, \tz-\tx +1)\,(2\, \tz-\tx )}
\eqn\PFthree
$$
degenerates to the two moduli system of the generic fiber, given by a
$K_3$ family of type $X_{12}(1,1,4,6)$. Actually, this $K_3$ is in
itself a elliptic fibration over $\IP^1$ with generic fiber
$X_6(1,2,3)$, as can be seen in an analogous way.

It is quite clear that elliptic fibrations lead very directly to
modular functions. Specifically, we present below a table of elliptic
curves ${\cal E}$, noticing that the present example corresponds to
the last entry.

{\vbox{\ninepoint{
$$
\vbox{\offinterlineskip\tabskip=0pt
\halign{\strut\vrule#
&\hfil~$#$
&\vrule#
&~~$#$~~\hfil
&~~$#$~~\hfil
&~~$#$~~\hfil
&~~$#$~~\hfil
&\vrule#\cr
\noalign{\hrule}
& &
& {\rm elliptic\ family }
&{\rm  PF\,\, operator}
& j(x)
& {\rm  mod.\ subgroup} &
\cr
\noalign{\hrule}
&{\cal E}_1 && X_{3}(1,1,1)&
\theta^2- 3 x ( 3 \theta+2)(3\theta+1)&
\displaystyle{(1+ 216x)^3\over x (1-27 x)^3} &
\Gamma(3)&
\cr
&{\cal E}_2 && X_{4}(1,1,2)&
\theta^2- 4 x ( 4 \theta+3)(4\theta+1)&
\displaystyle{(1+ 192x)^3\over x (1-64 x)^2}&
\Gamma(2)&\cr
&{\cal E}_3 && X_{6}(1,2,3)&
\theta^2- 12 x ( 6 \theta+5)(6\theta+1)&
\displaystyle{1\over x (1-432 x)} &
\Gamma^* &
\cr
\noalign{\hrule}}
\hrule}$$
\vskip-10pt
\noindent{\bf Table 1:} {\sl Families of elliptic curves ${\cal E}$,
their Picard-Fuchs operators, commensurability relations of the
mirror-map $x(q)$ as defined in \doubref\HKTYI\HKTYII\ with the
$j$-function, and the relevant modular subgroup of $SL(2,\ZZ)$. In
the first two cases the Hauptmodul of $\Gamma(3)$ and $\Gamma(2)$ is
related to the mirror map by removing from $1/x$ the constant term.
In the third case the commensurability polynomial is of genus one,
meaning that one needs two generators to define the function field on
${\cal E}_3$.} \vskip10pt}}}

The PF operator of ${\cal E}_3$ obviously represents the $y,z\to0$
limit of \PFthree. From the commensurability relation of the mirror
map of ${\cal E}_3$ with $j(q_x)$ we can immediately see that the
mirror map of \npolI\ in the limit $z,y\rightarrow 0$ is given by
$$
x(q_x)\ =\
{2\over j(q_x)+\sqrt{j(q_x)(j(q_x)-1728)}} \ .
\eqn\xqx
$$
In addition, it follows from analyzing the corresponding degeneration
limits of the PF-system that the mirror-maps $y(q_y)$ (and $z(q_z)$)
specialize to rational functions on the boundary of the moduli space,
$x=z=0$ ($x=y=0$, resp.): $y(q_y)={q_y\over (1+q_y)^2}$,
$z(q_z)={q_z\over (1 + q_z)^2}$. We will use the solution \xqx\ below
to provide further evidence in favor of the conjectured
heterotic-type II string duality.

Moreover, we can infer from Table 1 that the mirror map $x(q_x)$ (and
$y(q_y)$) of the hypersurface of bidegree $(3,3)$ in $\IP^1\times
\IP^1$, denoted $X_{3|3}(1,1,1|1,1,1)^{-162}_{2,83}$, of \HKTYI\ for
$y=0$ ($x=0$) is related to the Hauptmodul of $\Gamma(3)$. Finally,
we find that the mirror map\foot{The problem of including the twisted
sector in the analysis of the PF-system was recently solved in \BKK.}
$X_{12}(1,1,1,3,6)^{-324}_{3(1),165}$ is related to the Hauptmodul of
$\Gamma(2)$ at the boundary $y=z=0$.

\chapter{The three-moduli example $X_{24}(1,1,2,8,12)$ revisited}

We have seen how the appearance of elliptic functions in the mirror
maps of Calabi--Yau compactifications can naturally be understood in
terms of their special structure as fibrations, at least in the case
of one modulus, where we could use the results known in the
mathematical literature. Unfortunately, an analogous treatment for
more than one modulus does not seem to exist. We will now show that
the mirror map and Yukawa couplings of the three-moduli example of
\KV, $X_{24}(1,1,2,8,12)$, can nevertheless be written in terms of
elliptic functions in the weak coupling limit. This will provide a
further impressive non--trivial check on the conjecture of
equivalence of the corresponding \nex2 heterotic and type II strings.

\def\xb{\bar{x}} \def\yb{\bar{y}} \def\zb{\bar{z}}
\def\tone{\triangle_1}\def\ttwo{\triangle_2}
Following Kachru and Vafa \KV, we identify $y \sim e^{-\alpha S}
\rightarrow 0 $ with the weak coupling limit of the heterotic string
theory. This identification is motivated by the fact that the
discriminant locus of the mirror CY becomes a perfect square,
representing the splitting \SW\ of the classical $SU(2)$ singularity
into two branches in the quantum theory. Specifically, the
discriminant is
$$
\triangle = [\ (1-\zb)^2-\yb\ \zb^2\ ] \times
[\ ((1-\xb)^2-\xb^2\zb)^2 -\yb\ \xb^4\ ]
\times [\ 1-\yb\ ]
\equiv \tone \times \ttwo \times \triangle_3  \ ,
\eqn\dis
$$
where $\xb=432x, \yb=4y, \zb=4z$. For $\yb \rightarrow 0$,
$\triangle$ degenerates into quadratic factors that have the
following significance with respect to gauge symmetry enhancements in
the heterotic theory:
$$
\eqalign{
\tone = 0&: T=U\ \ \ \ \ \ \  SU(2) \cr
\ttwo = 0&: T=U=i \ \ SU(2)\times SU(2) \cr
 &\ \ \, T=U=\rho \ \ SU(3) \ .
}$$
The discriminant factor $\triangle_3$ has the interpretation
of a strong-coupling singularity in the heterotic theory.
The conjectured duality between the type II theory and the heterotic
theory implies that the perturbative $SO(2,2,\ZZ)$ symmetry of the
latter theory should be encoded in the former one. Indeed it turns
out that in the limit $\yb \rightarrow 0$ the mirror map for $\xb$
and $\zb$ can be written in terms of elliptic j-functions. More
precisely, using \xqx\ and the fact that $T$ and $U$ should enter
symmetrically, we find:
$$
\eqalign{
\xb &= q_1+\sum_{m+n>1} a_{mn} q_1^m\ q_3^n =
{\mu\over 2} {j(T)+j(U)-\mu\over j(T)j(U)
+\sqrt{j(T)(j(T)-\mu)}\sqrt{j(U)(j(U)-\mu)}} \cr
\zb &= q_3+\sum_{m+n>1} b_{mn}q_1^m\ q_3^n =
{(j(T)j(U) +\sqrt{j(T)(j(T)-\mu)}\sqrt{j(U)(j(U)-\mu)})^2\over
{j(T)j(U)(j(T)+j(U)-\mu)^2}}  , }
\eqn\mirmap
$$
where $\mu\equiv j(i)=1728$ and where we have defined $q_1\equiv
q_T$, $q_3\equiv q_U/q_T$\foot{For the definition of the
special coordinates $t_i=1/(2\pi i)\ln q_i$, see ref.\ \HKTYI.}
(we also have verified directly that this corresponds to
solutions of the PF equations).

Although on the first glance complicated, eqs.\ \mirmap\ can
be recognized as appropriate generalization of the various limits.
That is, for the $SU(2)$ enhanced line, $T=U$, we find
$$
\xb = {\mu\over 2j} \ , \qquad  \zb = 1  \ ,
$$
and consequently $\tone=0$. For the points of further enhancement we
get
$$
T=U=i\ :\ \xb = {1\over 2}\ , \ \ {\rm and}\ \qquad T=U=\rho\ :\ \xb
= \infty\, ,
$$
such that in addition $\ttwo = 0$. Moreover, in the limit
$U\rightarrow i \infty$ we recover \xqx:
$$
\zb = 0\ ,\qquad \xb = {\mu\over 2} {1\over
j(T)+\sqrt{j(T)(j(T)-\mu)}}
\eqn\mirlimIII
$$
(and similarly for $T\rightarrow i \infty$). This equation also
reflects the invariance of the defining polynomial \firstex \ under a
subgroup of general automorphisms: $x_i \rightarrow x_i$,
$i=3,4,5$, $x_2 \rightarrow x_2 + a(x_3x_4x_5)^2$, $x_1 \rightarrow
x_1 + b(x_3x_4x_5)^3+c x_2x_3x_4x_5$ that acts non-trivially on the
moduli space as follows:
$$
\psi_0 \rightarrow i\psi_0\ , \psi_1 \rightarrow \psi_1+2 \mu
\psi_0^6,
\psi_2 \rightarrow \psi_2 \ \
\eqn\symI
$$
and hence:
$$
\chi_1 \rightarrow {\chi_1\over \chi_1-1} \  \ (\ZZ_2)
$$
on $\chi_1 \equiv 1/\xb$. Identifying the invariant expression with
$j(T)$ reproduces \mirlimIII. Note also that the symmetry \symI\
exchanges the factors of the discriminant \dis:
$$
\tone \rightarrow \xb^4 \ttwo \ , \ \
\ttwo \rightarrow {1\over(1-\xb)^4} \ttwo \ .
$$

The identifications \mirmap\ provide a further, highly non--trivial
check on the conjectured string duality. For this purpose we need the
translation of the Yukawa couplings (that were determined in \HKTYI)
in terms of $S,T$ and $U$, as well as the expression for the mirror
map of the third Calabi--Yau modulus, $\yb$. It has the general form
$$
\yb = q_2 \sum_{m+n \geq 1}q^m_1 q^n_3 + {\cal O}(q_2^2)
\equiv q_s f_y(q_1,q_3) + {\cal O}(q_2^2) \ ,
\eqn\ybmirror
$$
with $q_s = e^{-8 \pi^2 S}$, where $S$ will be identified
with the (tree level) dilaton of the heterotic string.
Then, using
$$
\eqalign{
&\partial_T \xb = -j_T(T){\sqrt{j(U)(j(U)-\mu)}\over
\sqrt{j(T)(j(T)-\mu)}}
{1\over j(T)+j(U)-\mu} \xb (1-\xb) \cr
&\partial_T \yb = \yb \ \partial_T \ln f_y(q_1,q_3)  \cr
&\partial_T \zb = -j_T(T)\zb
\ \times\ \cr&
{\sqrt{j(T)(j(T)-\mu)}
(j(T)+j(U)-\mu)-2j(T)\sqrt{j(U)(j(U)-\mu)}(1-\xb)\over
j(T) \sqrt{j(T)(j(T)-\mu)}(j(T)+j(U)-\mu)}\cr
&\partial_S \xb = \partial_S \zb = 0\ , \ \ \partial_S \yb = -8\pi^2\
\yb \ ,
}$$
($j_T(T)\equiv\partial_T j(T)$) and the analogous relations obtained
by exchanging $T$ and $U$, the CY Yukawa couplings given in \HKTYI\
when written in terms of $S,\ T,\ U$ read:
\def\eft{E_4(T)}\def\est{E_6(T)}\def\efu{E_4(U)}\def\esu{E_6(U)}
\def\Kt{\tilde K}
$$\eqalign{
\Kt_{SSS}&=\Kt_{SST}=\Kt_{SSU}=\Kt_{STT}=\Kt_{SUU}=0 \cr
\Kt_{STU}&=  1 \cr
\Kt_{TTT} &= {i\over 2\pi}
{\eft\efu\esu(\eft^3-\est^2)\over \efu^3\est^2-\eft^3\esu^2}
\cr&
= -{1\over 4 \pi^2} {j_T(T)^{2} j(U)(j(U)-\mu)
\over (j(T)-j(U))j(T)(j(T)-\mu)j_U(U)} \cr
\Kt_{TTU} &=
-{i\over 2\pi} {\eft^2\est(\efu^3-\esu^2)
\over \efu^3\est^2-\eft^3\esu^2} + {i \over 2\pi}
\partial_T \ln f_y(q_1,q_3) \cr
&= {1 \over 4 \pi^2} {j_T(T)\over j(T)-j(U)}
+ {i \over 2\pi} \partial_T \ln f_y(q_1,q_3) \ .
}\eqn\crucialtest
$$
Here $E_{4,6}$ are the normalized Eisenstein series, and $\Kt_{ABC} =
1/(2 \pi^2 \omega_0)^2 K_{ABC}$, where
$\omega_0=E_4(T)^{1/4}E_4(U)^{1/4}$ is the fundamental period and the
transition from $K_{ABC}$ to $\Kt_{ABC}$ corresponds to going to the
canonical gauge. The expressions \crucialtest\ must be compared with
the results from perturbative string calculations performed in the
heterotic theory \doubref\dWKLL\AFGNT. We find indeed perfect
agreement~! Note also that corrections to $K_{TTT}$ arising from the
$T$-dependence of $\yb$ cancel in a very non--trivial way. This is
just as expected: for the coupling $K_{TTT}$ we do know the exact expression
from the calculations in the heterotic theory, while the corrections
to $K_{TTU}$ proportional to $f_y(q_1,q_3)$ are not known (this
coupling has been determined only to the leading order in
$T-U$).

As was pointed out to us \PrivComm, the mirror map
for $\yb$ provides a further independent consistency check. In fact,
since $\yb$ is invariant under the CY monodromy group that contains
the modular groups for $T$ and $U$, its logarithm should to be
identified with the modular invariant dilaton defined in \dWKLL:
$$
S_{inv} = S - {1\over 2} \partial_T\partial_U h^{(1)}(T,U) -
{1 \over 8\pi^2} \ln (j(T)-j(U)) + {\rm const.}\ .
\eqn\sinv
$$
Here $h^{(1)}(T,U)$ is the moduli dependent one-loop contribution to
the holomorphic prepotential. Since $h^{(1)} (T,U)$ transforms as
modular form of weight $(-2,-2)$ under $(T,U)$ duality
transformations up to terms quadratic in $T$ and $U$
\doubref\dWKLL\AFGNT, $\partial^3_T\partial^3_U h^{(1)}(T,U)$ is a
modular form of weight $(4,4)$. Consistency of \sinv\ and \ybmirror\
then implies
$$
\partial^2_T\partial^2_U \ln f_y(q_1,q_3) =
M(q_1,q_3) + \alpha\, \partial^2_T\partial^2_U \ln(j(T)-j(U))\ ,
\eqn\cons
$$
where $M$ is a modular form of weight $(4,4)$ that is regular in the
fundamental domain, except for a fourth order pole at $T=U$.
We indeed find that \cons \ is fulfilled, with $\alpha=1$ and $M=4\pi^2
\partial^3_U \Kt_{TTT}$.

\chapter{Duality symmetries}

Having successfully met further non-trivial checks on the
heterotic--type II duality for the model $X_{24}(1,1,2,8,12)$, we now
address some questions beyond leading perturbation theory. Adopting
the duality hypothesis, we know that important information about the
non--perturbative S- and T-duality of the heterotic string must be
encoded in the symmetries of the CY moduli space. Its monodromy
properties have been explicitly worked out \COFKM\ for the two-moduli
example $X_8(1,1,2,2,2)$, and can be straightforwardly determined for
other examples as well, using known results for the periods. Another
type of symmetries arise from the defining polynomial,
such as the automorphisms \symI. Alternatively, we may directly look
for symmetries in the instanton expansions, which reflect some of the
monodromy properties.

\noindent More specifically, analyzing various instanton expansions,
we are lead to consider the following symmetries acting on $q_i$~:

$$
\vbox{\offinterlineskip\tabskip=0pt
\halign{\strut
#
&\hfil~$#$
&{}~~$#$~~\hfil
&~~$#$~~\hfil
&~~$#$~~\hfil
&~~$#$~~\hfil
&~~$#$~~\hfil
&~~$#$~~\hfil
&~~$#$~~\hfil
&
#
\cr
& a)
&X_{24}(1,1,2,8,12)&
\
&q_1\rightarrow q_1q_3
&q_3 \rightarrow 1/q_3\ \ (q_2\equiv 0)
&\ \ \ \ \zb
& T \leftrightarrow U
&
&\cr

& b)
&\hfil\ \ "\ \hfil &
\
&q_3\rightarrow q_2q_3
&q_2 \rightarrow 1/q_2
&\ \ \ \ \yb
&\ \ \ ?
&
&\cr
&c)
&X_8(1,1,2,2,2)&
\
&q_1\rightarrow q_1q_2
&q_2 \rightarrow 1/q_2
&\ \ \ \ \yb
&\ \ \ ?
&
&
\cr
&\
&X_{12}(1,1,2,2,6)&
\
&q_1\rightarrow q_1q_2
&q_2 \rightarrow 1/q_2
&\ \ \ \ \yb
&\ \ \ ?
&
&\cr
}
}$$

\ni We indicated in the last two colomns the relevant modulus and the
interpretation of the symmetry in terms of the dual heterotic string;
evidence for what the question marks should stand for will be
presented below.

Symmetry {\sl a}) is easily identified as the mirror symmetry of the
heterotic string that exchanges the K\"ahler and complex structure
moduli of the compactification torus, $T_2$. We did not found a
simple generalization to $q_2\neq0$, possibly indicating a
non-perturbative breaking of this symmetry. Furthermore $\zb|_{q_3=1}\neq
1$ for $q_2\neq0$, that there is a shift of the singularity $T=U$.
However a clarification of these points is of course strongly
connected to the precise relation between heterotic and CY moduli at
all orders in $q_2$, an information which is beyond the present
knowledge.

A consequence of the inversion symmetry $q_3 \rightarrow 1/q_3$ is
that the mirror map has the property that powers of $q_3$ are
accompagnied by a sufficient number of powers of $q_1$. Translated
into the heterotic string language, this means that no negative
powers of $q_T,q_U$ appear in the expansions of physical quantities
such as the Yukawa couplings.\foot{More precisely, the mirror map for
$\zb$ has an additional overall factor of $q_3$ that is compatible
with \crucialtest.} In fact, this property can be traced back to the
special form of the basis vectors of the Mori cone. Furthermore, note
that the only association of $t_3$ with the moduli $T$, $U\in\IH^+$
of the heterotic string that is consistent with the symmetry
$q_3\rightarrow 1/q_3$, is given by: $q_3= q_T/q_U$.
Finally, remember from section 2 that the mirror map for $\zb$
reduces to $q_3/(1+q_3)^2$ in the limit $q_1,q_2 \rightarrow 0$.

A crucial observation is that these features are shared also by the
other two symmetries, {\sl b}) and {\sl c}). However, these
transformations are quite different from the point of view of the
heterotic string, in that they involve $q_2$ that is related to the
dilaton $S$.

If we allow $q_2$ to have additional dependence on $T$, and define
for convenience $q_2=q_S^\prime q_T^\alpha$, the symmetry {\sl c})
translates to $T\rightarrow(1+\alpha)T+S^\prime,\ S^\prime
\rightarrow -(1+\alpha) S^\prime-\alpha(2+\alpha)T$. Realizing the
shift symmetries $t\rightarrow t+1$ for $t=t_1,t_2,S^\prime,T$,
requires then $\alpha$ to be an integer. Imposing in addition the
reasonable condition that the positivity of the imaginary parts of
$S^\prime$ and $T$ (which are $\sim g^{-2}$ and $\sim R^2$,
respectively) is preserved, enforces $\alpha = -1$, and this implies:
$q_2=q_S/q_T$. That is, the natural interpretation of the symmetry
{\sl c}) is a symmetry under exchange of $S$
and $T$ (and similarly of $S$ and $U$ for {\sl b}))~! Note also that
a linear change of variables, $S^\prime \rightarrow S^\prime + \alpha
T$, does not alter the expressions for the Yukawa couplings, apart
from a constant shift.

Finally, we point out a remarkable property of the mirror maps
$\zb,\yb$ $(\yb)$ in the models $X_{24}\  (X_8,X_{12})$. That is,
$$
\zb_{X_{24}}|_{q_2=0 \atop q_3=1} = 1\ , \ \ \
\yb_{X_{24}}|_{q_2=1} = 1\ , \ \ \
\yb_{X_8,X_{12}}|_{q_2=1}= 1\ ,
$$
are independent of $q_1$ (and of $q_3$ for the second equation).
Comparing with the discriminants of these models, we see that the
singularities $\zb=1\ (cf., T=U)$, $\yb=1$ (cf., strong coupling) are
precisely at the fixed points of the symmetries {\sl a})--{\sl c}).
This means that the singularities correspond to $t_i=0$ of the
relevant moduli. Keeping in mind the relation $t_i=(\sum_i
A_i^j\omega_j) /\omega_0$, where $\omega_i$ are the (generically
independent) components of the CY period vector, we see that this
implies the vanishing of certain linear combinations of the periods.
Hence, according to the $N=2$ mass formula \BPSmass, Bogomolnyi states
with the appropriate quantum numbers should become massless at these
points in the moduli space (provided they exist). The fact that these
singularities are not of the simple conifold type indicates that the
ideas described in \Andy\ might apply in a more general context.

\chapter{Conclusions}

We have amassed further, and we think convincing, evidence in favor
of the conjectured duality between heterotic strings compactified on
$K_3\times T_2$, and type II strings compactified on certain
Calabi-Yau manifolds. We have also gained insight in the modular
properties of certain CY theories in the weak coupling limit, and
believe that by focusing on $K_3$ fibrations, many more examples can
be systematically studied. We also analyzed the symmetry
structure of some models, linking the symmetries of the CY
instanton expansion to the perturbative and non-perturbative $T$- and
$S$-dualities of the quantum heterotic string. We hope that the
methods described in this paper can be developed further, allowing to
make new, concrete and truly non-perturbative predictions for \nex2
supersymmetric string theory.

\ack

We like to thank P.\ Aspinwall, S.\ Katz, B.\ Lian, C.\ Vafa,
S.T.~Yau, and especially S.\ Theisen for discussions.

\app{Calabi-Yau manifolds that are $K_3$ fibrations}

$$
\vbox{\offinterlineskip\tabskip=0pt
\halign{
\strut\vrule#
&\hfil~$#$
&\vrule# &
\hfil~~$#$~~&
\hfil~~$#$~~&
\hfil~~$#$~~
&\vrule# &
\hfil~~$#$~~&
\hfil~~$#$~~
&\vrule# \cr
\noalign{\hrule}
& && h_{1,1}&h_{2,1}&\chi&& {\rm deg.}& {\rm weights}&\cr
\noalign{\hrule}
&1&&     2&    86&  -168&&     8& (    1,    1,    2,    2,    2)&\cr
&2&&     2&   128&  -252&&    12& (    1,    1,    2,    2,    6)&\cr
&3&&     3&    99&  -192&&    10& (    1,    1,    2,    2,    4)&\cr
&4&&     3&   243&  -480&&    24& (    1,    1,    2,    8,   12)&\cr
&5&&     4&   148&  -288&&    16& (    1,    1,    2,    4,    8)&\cr
&6&&     4&   190&  -372&&    20& (    1,    1,    2,    6,   10)&\cr
&7&&     5&   101&  -192&&    12& (    1,    1,    2,    4,    4)&\cr
&8&&     5&   121&  -232&&    14& (    1,    1,    2,    4,    6)&\cr
&9&&     5&   161&  -312&&    18& (    1,    1,    2,    6,    8)&\cr
&10&&     7&   143&  -272&&   20& (    1,    1,    4,    4,   10)&\cr
&11&&     7&   271&  -528&&   36& (    1,    1,    4,   12,   18)&\cr
&12&&     8&   104&  -192&&   16& (    1,    1,    4,    4,    6)&\cr
&13&&     8&   164&  -312&&   24& (    1,    1,    4,    6,   12)&\cr
&14&&     8&   194&  -372&&   28& (    1,    1,    4,    8,   14)&\cr
&15&&     9&   111&  -204&&   18& (    1,    1,    4,    6,    6)&\cr
&16&&     9&   125&  -232&&   20& (    1,    1,    4,    6,    8)&\cr
&17&&     9&   153&  -288&&   24& (    1,    1,    4,    8,   10)&\cr
&18&&     9&   321&  -624&&   48& (    1,    1,    6,   16,   24)&\cr
&19&&    10&   194&  -368&&   32& (    1,    1,    6,    8,   16)&\cr
&20&&    10&   220&  -420&&   36& (    1,    1,    6,   10,   18)&\cr
&21&&    10&   376&  -732&&   60& (    1,    1,    8,   20,   30)&\cr
&22&&    11&   131&  -240&&   24& (    1,    1,    6,    8,    8)&\cr
&23&&    11&   143&  -264&&   26& (    1,    1,    6,    8,   10)&\cr
&24&&    11&   167&  -312&&   30& (    1,    1,    6,   10,   12)&\cr
&25&&    11&   227&  -432&&   40& (    1,    1,    8,   10,   20)&\cr
&26&&    11&   251&  -480&&   44& (    1,    1,    8,   12,   22)&\cr
&27&&    11&   485&  -960&&   84& (    1,    1,   12,   28,   42)&\cr
&28&&    12&   164&  -304&&   32& (    1,    1,    8,   10,   12)&\cr
&29&&    12&   186&  -348&&   36& (    1,    1,    8,   12,   14)&\cr
&30&&    12&   318&  -612&&   60& (    1,    1,   12,   16,   30)&\cr
&31&&    13&   229&  -432&&   48& (    1,    1,   12,   16,   18)&\cr
\noalign{\hrule}}
\hrule}$$
\noindent{\bf Table A.1:} {\sl Simple hypersurfaces in weighted
$\IP^4$ which are $K_3$ fibrations. } \vskip20pt

$$\vbox{\offinterlineskip\tabskip=0pt
\halign{
\strut\vrule#
&\hfil~$#$
&\vrule# &
\hfil~~$#$~~&
\hfil~~$#$~~
&\vrule# &
\hfil~~$#$~~&
\hfil~~$#$~~&
\hfil~~$#$~~
&\vrule# \cr
\noalign{\hrule}
& && h_{1,1}&h_{2,1}&& d_1&d_2& {\rm weights}& \cr
\noalign{\hrule}
&1&&     2&    68&&     6&     4&  (  1,    1,    2,    2,    2,    2)&\cr
&2&&     3&    69&&     6&     6&  ( 1,    1,    2,    2,    2,    4)&\cr
&3&&     4&    84&&     8&     8&  (  1,    1,    2,    2,    4,    6)&\cr
&4&&     4&    76&&     8&     6&  (  1,    1,    2,    2,    4,    4)&\cr
&5&&     6&   138&&    16&    12&  (  1,    1,    2,    6,    8,   10)&\cr
&6&&     6&   102&&    12&    10&  (  1,    1,    2,    4,    6,    8)&\cr
&7&&     6&    98&&    12&     8&  ( 1,    1,    2,    4,    6,    6)&\cr
&8&&     6&    82&&    10&     8&  (  1,    1,    2,    4,    4,    6)&\cr
&9&&     6&    70&&     8&     8&  (  1,    1,    2,    4,    4,    4)&\cr
&10&&     8&    76&&    12&     8& (   1,    1,    4,    4,    4,    6)&\cr
&11&&     9&    81&&    12&    12& (   1,    1,    4,    4,    6,    8)&\cr
&12&&     9&    75&&    12&    10& (   1,    1,    4,    4,    6,    6)&\cr
&13&&    10&   122&&    20&    16& (   1,    1,    4,    8,   10,   12)&\cr
&14&&    10&    98&&    16&    14& (  1,    1,    4,    6,    8,   10)&\cr
&15&&    10&    94&&    16&    12& (   1,    1,    4,    6,    8,    8)&\cr
&16&&    10&    84&&    14&    12& (   1,    1,    4,    6,    6,    8)&\cr
&17&&    10&    76&&    12&    12& (  1,    1,    4,    6,    6,    6)&\cr
&18&&    12&   128&&    24&    20&  (  1,    1,    6,   10,   12,   14)&\cr
&19&&    12&   108&&    20&    18&  ( 1,    1,    6,    8,   10,   12)&\cr
&20&&    12&   104&&    20&    16&  (  1,    1,    6,    8,   10,   10)&\cr
&21&&    12&    96&&    18&    16&  (  1,    1,    6,    8,    8,   10)&\cr
&22&&    13&   139&&    28&    24&  (  1,    1,    8,   12,   14,   16)&\cr
&23&&    13&   121&&    24&    22&  (  1,    1,    8,   10,   12,   14)&\cr
&24&&    13&   117&&    24&    20&  (  1,    1,    8,   10,   12,   12)&\cr
&25&&    14&   166&&    36&    32&  (  1,    1,   12,   16,   18,   20)&\cr
\noalign{\hrule}}
\hrule}$$
\noindent{\bf Table A.2:} {\sl Simple complete intersections
Calabi-Yau spaces in weighted
$\IP^5$ which are $K_3$ fibrations. }

\listrefs
\end